\documentclass[twocolumn,showpacs,aps,epsfig,nofootinbib]{revtex4}

\usepackage{graphicx}
\usepackage{epstopdf}
\usepackage{latexsym}
\usepackage{amssymb}
\usepackage{braket}
\usepackage{color}

\usepackage[center]{subfigure}

\begin{document}

 \newcommand{\bq}{\begin{equation}}
 \newcommand{\eq}{\end{equation}}
 \newcommand{\bqn}{\begin{eqnarray}}
 \newcommand{\eqn}{\end{eqnarray}}
 \newcommand{\nb}{\nonumber}
 \newcommand{\lb}{\label}
\newcommand{\PRL}{Phys. Rev. Lett.}
\newcommand{\PL}{Phys. Lett.}
\newcommand{\PR}{Phys. Rev.}
\newcommand{\CQG}{Class. Quantum Grav.}

\title{Inflation in general covariant Ho\v{r}ava-Lifshitz gravity without projectability}

\author{Tao Zhu   ${} ^{a}$}
\email{zhut05@gmail.com}

\author{Yongqing Huang ${} ^{a, b}$}
\email{yongqing_huang@baylor.edu}

\author{Anzhong Wang ${} ^{a,b}$ \footnote{The corresponding author}}
\email{anzhong_wang@baylor.edu}

\affiliation{${} ^{a}$ Institute for Advanced Physics $\&$ Mathematics, Zhejiang University of Technology, Hangzhou 310032,  China\\
${} ^{b}$GCAP-CASPER, Physics Department, Baylor University, Waco, TX 76798-7316, USA}

\date{\today}

\begin{abstract}
In this paper, we study inflation in the general covariant Ho\v{r}ava-Lifshitz gravity without the projectability condition.  We write down explicitly the equations of the linear scalar perturbations of the  FRW universe for a single scalar field without specifying to any gauge.  Applying these equations to a particular gauge, we are able to obtain a master equation of the perturbations, in contrast to all the other versions of the theory without the projectability condition. This is because in the current version of the theory  it has the same degree of freedom of general relativity. To calculate the power spectrum and index, we first define the initial conditions as the ones that  minimize the energy of the ground state. Then, we  obtain the full solutions of the equation of motion, by using the WKB approximations. From these solutions, we calculate the power spectrum and spectrum index of the comoving curvature perturbations and  find the corrections due to the high order spatial derivative terms of the theory to those standard results obtained in general relativity. Remarkably, partly of corrections are a direct consequence  of the non-projectability condition. It is also shown that the perturbations are still of scale-invariance, and   the results obtained in  the general covariant Ho\v{r}ava-Lifshitz gravity without the projectability condition are consistent with all current cosmological observations.
\end{abstract}


\maketitle

\section{Introduction}
\renewcommand{\theequation}{1.\arabic{equation}} \setcounter{equation}{0}

Recently, Ho\v{r}ava formulated a theory of quantum gravity, whose scaling at short distances exhibits a strong anisotropy between space and time \cite{Horava},
\bq
\lb{1.1}
{\bf x} \rightarrow b^{-1} {\bf x}, \;\;\;  t \rightarrow b^{-z} t.
\eq
In order for the theory to be power-counting renormalizable, in $(3+1)$-dimensions the critical exponent $z$  needs to be $z \ge 3$ \cite{Horava,Visser}. The gauge symmetry of the theory now is broken from the general covariance, $\delta{x}^{\mu} = - \delta{x}^{\mu}(t, x)\; (\mu = 0, 1, 2, 3),$ down to the foliation-preserving diffeomorphisms Diff($M, \; {\cal{F}}$),
\bq
\lb{1.4}
\delta{t} = - f(t),\; \;\; \delta{x}^{i}  =   - \zeta^{i}(t, {\bf x}).
\eq
Abandoning  the general covariance   gives rise to a proliferation of independently coupling constants \cite{BPS,KP}, which could potentially limit   the prediction powers of the theory. To reduce the number of these constants, Ho\v{r}ava imposed two conditions, {\em the projectability and detailed balance} \cite{Horava}. The
former assumes that the lapse function $N$ in the  Arnowitt-Deser-Misner  (ADM) decompositions  \cite{ADM} is a function of $t$ only,
\bq
\lb{1.6}
N = N(t),
\eq
while the latter assumes that  gravitational potential ${\cal{L}}_{V}$ can be obtained from a superpotential $W_{g}$ that is defined on the three-spatial hypersurfaces $t = $ Constant. With these conditions, the number of independently coupling constants reduces to $5$ \cite{Horava}. 
 
However, with the detailed balance  condition, the Newtonian  limit does not exist \cite{LMP}, and  a scalar field in the UV is not stable  \cite{CalA}. Thus, it is generally believed that this condition should be abandoned  \cite{KK}.  But,  due to    several remarkable features \cite{Hreview}, Borzou, Lin and Wang  recently studied it in detail, and found that   the scalar field  can be stabilized in both regimes, UV and IR, if the detailed balance  condition is allowed to be broken softly \cite{BLW}. 
 
On the other hand, giving up the  detailed balance condition, but still keeping the projectability one, the number of the independent coupling  constants can also be significantly reduced. In fact, together with the assumptions of the parity and time-reflection symmetry, it can be reduced from more than 70 to 11  \cite{SVW} (See also \cite{KKa}).
However, in this model the Minkowski spacetime is  not stable \cite{SVW,WM} \footnote{In the literature, the ghost problem was often mentioned \cite{reviews,Mukc}. But, by restricting the
coupling constant $\lambda$ to the regions $\lambda \ge 1$ or $\lambda < 1/3$, this problem is solved (at least in the classical level) \cite{Horava,SVW,BS,WM}.
 In addtion, when $\lambda
\in(1/3, 1)$, the instability problem disappears. Therefore, one of these two problems can  be always avoided by properly choosing $\lambda$.
In this paper, we choose  $\lambda \ge 1$, so the  ghost problem does not exist.}, 
although the de Sitter spacetime is \cite{HWW,WWa}.  In addition, such a theory also faces
the strong coupling problem \cite{SC,WWa} \footnote{It should be noted that strong coupling is not a real problem, but an indication that the linear perturbations
involved are broken down, and nonlinear effects are needed to be taken into account. If the theory is consistent with observations, after those nonlinear effects are taken into 
account, it still represents a viable theory. One example is the massive gravity \cite{Vain}.  In the content of the Ho\v{r}ava-Lifshitz theory with the projectability condition,
such effects were also studied in  both  spherically symmetric static
spacetimes \cite{Mukc} and cosmological models \cite{Izumi:2011eh,GMW,WWa}, and found that   the spin-0 gravitons decouple after nonlinear effects are taken into account. 
As a result, the relativistic limits indeed exist.}.  Note that both of these two problems are closely related to the existence of a spin-0
graviton \cite{reviews,Mukc}, because of the   foliation-preserving diffeomorphisms \footnote{Since  the   foliation-preserving diffeomorphisms (\ref{1.4})  is also
assumed in  the version  without the projectability condition \cite{BPS}, the spin-0 graviton exists there, too.}. Another problem related to  the presence of this spin-0 graviton is
the difference of its speed from that of   the spin-2 graviton.  Since   they are not related by any symmetry, it  poses
a great challenge for any attempt to restore Lorentz symmetry  at low energies where it has been  well tested experimentally.  In particular, one needs a mechanism
to ensure that in those energy scales all species  of matter and gravity  have the same effective speed and light cones.

To overcome the above mentioned  problems, recently,  together with Shu and Wu,  two of the present authors proposed a model without the projectability condition, but
 assuming that:  (a) the detailed 
balance condition is softly broken; and (b) the symmetry of theory is enlarged to included a local U(1) symmetry \cite{ZWWS,ZSWW}
\footnote{The enlarged symmetry was first introduced  by Ho\v{r}ava and Melby-Thompson (HMT)
in  the case with the projectability condition and $\lambda = 1$ \cite{HMT}, and was soon generalized to the case with any $\lambda$  \cite{Silva}, where
$\lambda$ is a coupling constant, which characterizes the deviation from general relativity (GR) in the IR.  In such a setup, the spin-0 gravitons are eliminated, and the degree of the freedom
of the gravitional sector is the same as that in GR \cite{HMT,WW,Silva,HW,LWWZ,Kluson2}. In this model, cosmology and spherically symmetric spacetimes were 
also studied in, respectively, \cite{HWWb} and \cite{BSW,AP,LSW}.}, 
\bq
\lb{symmetry}
 U(1) \ltimes {\mbox{Diff}}(M, \; {\cal{F}}).
\eq
The detailed balance condition  considerably reduces the number of independently coupling constants, in addition to the desired features mentioned above, while allowing it to be softly breaking
yields a healthy IR limit. On the other hand, the presence of the local U(1) symmetry eliminates the existence of the spin-0 graviton, whereby all the problems related to it, including 
instability, strong coupling, and different speeds, are now all out of questions \cite{ZWWS,ZSWW}. The enlarged symmetry (\ref{symmetry}) is realized by intruding a U(1) gauge field
$A$, and a Newtonian prepotential $\varphi$, the same as those in the case with the projectability condition \cite{HMT}, where under the U(1) transformations, $A$ and $\varphi$ transform as,
\bq
\lb{uone}
\delta_{\alpha} A = \dot{\alpha} - N^{i}\nabla_{i}\alpha,\;\;\;
\delta_{\alpha}\varphi = - \alpha,
\eq
 where $\alpha$ denotes the U(1) generator,  $\dot{\alpha} \equiv d\alpha/dt$, $N^i$ is the shift vector in the ADM decompositions,
and $\nabla_{i}$ the covariant derivative with respet to the 3-metric $g_{ij}$, defined on the hypersurfaces $t= $ Constant. A remarkable feature of this model is that the lapse function 
$N$ depends on both $t$ and $x$, and the model has the same freedom (only  spin-2 gravitons exist) as that in GR. 
 
The Ho\v{r}ava-Lifshitz (HL) theory modifies GR in strong gravitational regimes. Therefore, a natural application of it is the early universe. In this paper,  we shall study inflation of a single scalar field
 in the setup of  \cite{ZWWS,ZSWW}, 
by paying particular attention on two important issues: (i) the consistency of the theory with cosmological observations;
and (ii) its distinguishable signatures  from other theories of gravity, including GR. Specifically, the paper is organized as follows. In Sec. II, we present a brief review of the general covariant Ho\v{r}ava-Lifshitz gravity without the projectability condition, while in Sec. III, we study the linear scalar perturbations, and discuss their  gauge choices and gauge-invariants. After working out explicitly the general equations for the cosmological linear scalar perturbations in a flat FRW universe without specifying to a particular gauge,   we derive the master equation (\ref{eom}) for the scalar perturbations in the specific gauge (\ref{gauge}). In Sec. IV,  we first define the initial vacuum state of the scalar perturbation as the one that   minimizes the energy of the ground state, and then calculate the power spectrum and index of the comoving curvature perturbation. Finally, in Sec. V we present our main conclusions.

It should be noted that linear perturbations of the Friedmann-Robertson-Walker (FRW) universe in the non-projectable HL theory without the enlarged symmetry   (\ref{symmetry}) has been studied by several authors \cite{Cosmo}.  Due to the presence of the spin-0 graviton modes, they are quite different from the ones presented here. In particular, because of the coupling of this mode to matter fields, 
 a master equation does not exist \cite{BF02}. 
 
 In addition,   in the current setup, the coupling constant $\Lambda_g$,  defined explicitly in Eq.(\ref{2.5}) below, has to vanish, 
although quantum mechanically it is expected to be  subjected to radiative corrections. It   is still an open question whether 
$\Lambda_{g} = 0$  is the low energy fixed pointed or not, as the  corresponding  RG flow of the theory  has not been worked 
out, yet. 

\section{General covariant Ho\v{r}ava-Lifshitz gravity without projectability}

\renewcommand{\theequation}{2.\arabic{equation}} \setcounter{equation}{0}

In this section, we shall give a very brief introduction to the general covariant HL  gravity with the enlarged symmetry (\ref{symmetry}) but
without the projectability condition. For detail, we refer readers to \cite{ZWWS,ZSWW}.
The total action of the theory can be written as,
\bqn 
\lb{TA}
S &=& \zeta^2\int dt d^{3}x  \sqrt{g}N \Big({\cal{L}}_{K} -
{\cal{L}}_{{V}} +  {\cal{L}}_{{A}}+ {\cal{L}}_{{\varphi}}  + \frac{1}{\zeta^2} {\cal{L}}_{M}\Big), \nb\\
\eqn
where $g={\rm det}(g_{ij})$, and
\bqn \lb{2.5}
{\cal{L}}_{K} &=& K_{ij}K^{ij} -   \lambda K^{2},\nb\\
{\cal{L}}_{A} &=&\frac{A}{N}\Big(2\Lambda_{g} - R\Big), \nb\\
{\cal{L}}_{\varphi} &=&  \varphi{\cal{G}}^{ij}\big(2K_{ij}+\nabla_i\nabla_j\varphi+a_i\nabla_j\varphi\big)\nb\\
& & +(1-\lambda)\Big[\big(\Delta\varphi+a_i\nabla^i\varphi\big)^2  
+2\big(\Delta\varphi+a_i\nabla^i\varphi\big)K\Big]\nb\\
& & +\frac{1}{3}\hat{\cal G}^{ijlk}\Big[4\left(\nabla_{i}\nabla_{j}\varphi\right) a_{(k}\nabla_{l)}\varphi \nb\\
&&  ~~ + 5 \left(a_{(i}\nabla_{j)}\varphi\right) a_{(k}\nabla_{l)}\varphi + 2 \left(\nabla_{(i}\varphi\right)a_{j)(k}\nabla_{l)}\varphi \nb\\
&&
~~ 
+ 6K_{ij} a_{(l}\nabla_{k)}\varphi 
\Big].
\eqn
Here $\Delta \equiv g^{ij}\nabla_{i}\nabla_{j}$,  and $\Lambda_{g}$ is a coupling constant. 
The Ricci and Riemann tensors $R_{ij}$ and $R^{i}_{\;\; jkl}$  all refer to the 3-metric $g_{ij}$, with $R_{ij} = R^{k}_{\;\;ikj}$ and
\bqn \lb{2.6}
K_{ij} &\equiv& \frac{1}{2N}\left(- \dot{g}_{ij} + \nabla_{i}N_{j} +
\nabla_{j}N_{i}\right),\nb\\
{\cal{G}}_{ij} &\equiv& R_{ij} - \frac{1}{2}g_{ij}R + \Lambda_{g} g_{ij}.
\eqn
${\cal{L}}_M$ is the Lagrangian of matter fields.

When the projectability condition is abandoned, it gives rise to 
a proliferation of a large number of independently coupling constant \cite{KP,BPS}. Following Horava, the detailed balance condition is generalized  to
\cite{ZWWS,ZSWW},
\bq
\lb{gdbc}
{\cal{L}}_{(V,D)} = \Big(E_{ij} \;\; A_{i}\Big)\left(\matrix{{\cal{G}}^{ijkl} & 0\cr 0 & - g^{ij}\cr}\right)\left(\matrix{E_{kl}\cr A_{j}\cr}\right),
\eq
where ${\cal{G}}^{ijkl}$ denotes the generalized De Witt metric, defined as ${\cal{G}}^{ijkl} = \frac{1}{2} \big(g^{ik}g^{jl} + g^{il}g^{jk}\big) - \lambda g^{ij}g^{kl}$,
and $E_{ij}$ and $A_{i}$ are given by 
\bq
\lb{gdbc1}
E^{ij} = \frac{1}{\sqrt{g}}\frac{\delta{W}_{g}}{\delta{g}_{ij}},\;\;\;
A^i = \frac{1}{\sqrt{g}}\frac{\delta W_a}{\delta a_i}.
\eq
The super-potentials $W_g$ and $W_a$ are constructed as 
\bqn
\lb{gdbc2}
W_{g}  &=& \frac{1}{w^{2}}\int_{\Sigma}{\omega_{3}(\Gamma)} + \mu \int{d^3x \sqrt{g}\Big(R - 2\Lambda\Big)},\nb\\
W_{a} &=& \int{d^{3}x \sqrt{g} \sum_{n= 0}^{1}{{\cal{B}}_{n}a^{i}\Delta^{n}{a_{i}}}},
\eqn
where $\omega_{3}(\Gamma)$ denotes the gravitational 3-dimensional   Chern-Simons term, $w,\; \mu,\; \Lambda$ and ${\cal{B}}_{n}$ are arbitrary  constants. However, to have a healthy 
infrared limit,   the detailed balance condition is allowed to be broken softly, by adding all the low dimensional operators, so that the potential finally takes the form
\cite{ZWWS,ZSWW},
\bqn 
\lb{2.5a}
 {\cal{L}}_{V} &=&  \gamma_{0}\zeta^{2}  -  \Big(\beta_0  a_{i}a^{i}- \gamma_1R\Big)
+ \frac{1}{\zeta^{2}} \Big(\gamma_{2}R^{2} +  \gamma_{3}  R_{ij}R^{ij}\Big)\nb\\
& & + \frac{1}{\zeta^{2}}\Bigg[\beta_{1} \left(a_{i}a^{i}\right)^{2} + \beta_{2} \left(a^{i}_{\;\;i}\right)^{2}
+ \beta_{3} \left(a_{i}a^{i}\right)a^{j}_{\;\;j} \nb\\
& & + \beta_{4} a^{ij}a_{ij} + \beta_{5}
\left(a_{i}a^{i}\right)R + \beta_{6} a_{i}a_{j}R^{ij} + \beta_{7} Ra^{i}_{\;\;i}\Bigg]\nb\\
& &    
 +  \frac{1}{\zeta^{4}}\Bigg[\gamma_{5}C_{ij}C^{ij}  + \beta_{8} \left(\Delta{a^{i}}\right)^{2}\Bigg],
 \eqn
where all the coefficients, $ \beta_{n}$ and $\gamma_{n}$, are
dimensionless and arbitrary, except for the ones of the sixth-order derivative terms,  $\gamma_{5}$ and $\beta_{8}$, which
must satisfy the conditions, 
\bq
\lb{2.8a}
 \gamma_{5} > 0, \;\;\; \beta_{8} <  0,
 \eq
in order for the theory to be unitary in the UV.  $C_{ij}$ denotes the Cotton tensor, defined by
\bq
\lb{1.12}
C^{ij} = \frac{ {{e}}^{ikl}}{\sqrt{g}} \nabla_{k}\Big(R^{j}_{l} - \frac{1}{4}R\delta^{j}_{l}\Big),
\eq
with  $e^{123} = 1$.
Using the Bianchi identities, 
one can show that $C_{ij}C^{ij}$ can be written in terms of the
five independent sixth-order derivative terms in the form
\bqn
\lb{1.13}
C_{ij}C^{ij}  &=& \frac{1}{2}R^{3} - \frac{5}{2}RR_{ij}R^{ij} + 3 R^{i}_{j}R^{j}_{k}R^{k}_{i}  +\frac{3}{8}R\Delta R\nb\\
& &  +
\left(\nabla_{i}R_{jk}\right) \left(\nabla^{i}R^{jk}\right) +   \nabla_{k} G^{k},
\eqn
where
\lb{1.14}
\bqn
G^{k}=\frac{1}{2} R^{jk} \nabla_j R - R_{ij} \nabla^j R^{ik}-\frac{3}{8}R\nabla^k R.
\eqn

To be consistent with observations in the IR, we assume that
\bq
\lb{2.8}
\zeta^{2} = \frac{1}{16\pi G},\;\;\;  \gamma_{1} = -1,
\eq
where $G$ denotes the Newtonian constant, and
\bq
\lb{2.8b}
\Lambda \equiv \frac{1}{2} \zeta^{2}\gamma_{0},
\eq
is the cosmological constant. 

 It should be noted that it is still not clear whether the detailed balance condition remains 
after quantum effects are taken into account \cite{Horava,reviews}, despite the fact that
 it has several desirable features \cite{BLW,Hreview}, and helps  to reduce the number of independent constants significantly. Without such a condition, the total number of
 independent coupling constants are about 100 \cite{KP,BPS,ZWWS}.

Variations of the total action (\ref{TA}) with respect to $N, \; N^{i}, \; A, \; \varphi$ and $g_{ij}$ yield, respectively, the Hamiltonian, momentum, $A$-, and $\varphi$-constraints, and dynamical equations,
which are given explicitly in \cite{ZSWW}. From those equations, one can see clearly that the mathematical  structure of the theory is completely different from the one without the U(1) symmetry \cite{BPS}.

In addition, assuming the translation symmetry of the action (\ref{symmetry}), one obtains the conservation laws of energy and momentum \cite{ZSWW}. However, 
because of the reduced symmetry, $\delta{t} = f(t)$, the energy is conserved only globally, in contrast to that of GR.

\section{Inflation of a single scalar field}
\renewcommand{\theequation}{3.\arabic{equation}} \setcounter{equation}{0}

Inflation in the projectable case with
the enlarged symmetry (\ref{symmetry}) has been studied recently in \cite{HWWb}, and in this section we shall closely follow it, although the two setups
have completely  different mathematical as well as physical structures, as to be shown below. 

A scalar field  in the current setup is described by 
\cite{BLW,WWM,ZSWW},
\bqn\lb{calAa}
S_\chi=\int dt d^3x \sqrt{g}N \mathcal{L}_M,
\eqn
where
\bqn
\lb{chiactionA}
{\cal{L}}_M &=& {\cal{L}}_\chi^{(A,\varphi)}+{\cal{L}}_\chi^{(0)},\nb\\
{\cal{L}}_\chi^{(A,\varphi)} &=& \frac{A-{\cal{A}}}{N}\left[c_1(\chi) \Delta \chi + c_2(\chi) (\nabla \chi)^2\right] \nb\\
&&+ \frac{f}{2} [(\nabla^k \varphi)(\nabla_k \chi)]^2 \nb\\
&& - \frac{f}{N} (\dot{\chi} - N^i \nabla_i \chi) (\nabla^k \varphi) (\nabla_k \chi) ,
\\
\lb{chiactionB}
{\cal{L}}_\chi^{(0)} &=& \frac{f}{2N^2} (\dot{\chi}-N^i \nabla_i \chi)^2 - {\cal{V}},
\eqn
and
\bqn
\lb{chiactionC}
{\cal{V}} &=& V(\chi) + \left(\frac{1}{2}+V_1(\chi)\right)(\nabla \chi)^2 + V_2(\chi) {\cal{P}}^2_1  \nb\\
&&+ V_3(\chi) {\cal{P}}^3_1 + V_4(\chi) {\cal{P}}_2 + V_5(\chi) (\nabla \chi)^2 {\cal{P}}_2 \nb\\
&&+ V_6 {\cal{P}}_1 {\cal{P}}_2,\\
\lb{calA}
{\cal{A}}&\equiv& - \dot{\varphi} + N^i \nabla_i \varphi + \frac{1}{2} N (\nabla_i \varphi) (\nabla^i \varphi),\nb\\
{\cal{P}}_n &\equiv& \Delta^n \chi, \;\;\;\;\;\;\; V_6 \equiv - \sigma_3^2,
\eqn
where $\sigma_{3}$ is a constant. The coefficient $f$ in (\ref{calAa}) is a function of $\lambda$ only, subjected to the requirements:
 (i) the scalar field must be ghost-free in all the energy scales, including the UV and IR; (ii) in the IR limit, the scalar field has a well-defined velocity, 
 which should be equal or very closed to its relativistic value; and (iii) the stability condition in the IR requires $f(\lambda)>0$ \cite{BLW,ZSWW}. 
 Usually, one chooses $f = 1$. In this paper, we shall leave this possibility open. 
 To study the inflationary model of such a scalar field, we first consider the slow-roll conditions in the FRW background.

\subsection{Slow-Roll inflation}

We consider a general FRW space-time,
\bqn
ds^2=a^2(\eta)(-d\eta^2+\gamma_{ij} dx^idx^j),
\eqn
where
\bqn
\gamma_{ij}=\frac{\delta_{ij}}{(1+kr^2/4)^2},
\eqn
and $k = 0, \pm 1$. Then, it can be shown that  the $A$- and $\varphi$- constraints yield \cite{ZSWW},
\bqn
\lb{eq6a}
\label{var}\frac{{\cal{H}} }{a} \left(\Lambda_g - \frac{k}{a^2}\right) = - \frac{8 \pi G}{3} \hat{J}_\varphi,\\
\lb{eq6b}
\label{bA}\frac{k}{a^2} - \frac{\Lambda_g}{3} = \frac{4\pi G \hat{J}_A}{3},
\eqn
where ${\cal{H}} = (da/d\eta)/a \equiv a'/a$, and 
\bqn
\hat{J}_\varphi &\equiv& -\frac{\delta {\cal{L}}_M}{\delta \varphi} =0,\nb\\
%
 \hat{J}_A &\equiv& 2 \frac{\delta ( N {\cal{L}}_M)}{\delta A} = 0,
\eqn
as one can see from Eqs.(\ref{chiactionA}) - (\ref{calA}) when only the FRW background is considered. 
Then, from Eqs.(\ref{eq6a}) and (\ref{eq6b}) we find that
 \bq
 \lb{KL}
 k=0=\Lambda_g. 
 \eq
That is, the FRW universe is necessarily flat  in the current setup for the scalar field $\chi$,
given by Eq.(\ref{calAa}). It is interesting to note that in such a setup, the coupling constant $\Lambda_g$ has to vanish, 
although quantum mechanically it is expected to be  subjected to radiative corrections. It   is still an open question whether 
$\Lambda_{g} = 0$ is the low energy fixed pointed or not, as the  corresponding  RG flow of the theory  has not been worked 
out, yet.  Since we assume that inflation occurred 
in the regime $H \ll M_{*}$ [See the discussions given below], where
$H$ is the Hubble constant during the inflation epoch, and $M_{*}$ the suppression energy scale of the high-order derivative terms
\cite{ZSWW}, in the rest of the paper  it is sufficient  for us to consider that Eq.(\ref{KL}) holds classically. 
  
  On the other hand, from the dynamical equations, we find that
 \bqn
  \lb{fried1}
{\cal{H}}^2 &=& \frac{8\pi \tilde{G}a^2}{3} \left(\frac{1}{2}\hat{\chi}'^2+\tilde{V}(\hat{\chi})\right),\\ 
\lb{fried2}
2{\cal{H}}'+{\cal{H}}^2 &=& 8\pi G a^2 \left(\frac{1}{2}\hat{\chi}'^2-\tilde{V}(\hat{\chi})\right), 
  \eqn
 where $\hat{\chi} = \hat{\chi}(\eta)$ denotes the background scalar field, and 
 \bq
 \lb{5.13c}
 \tilde{G} \equiv \frac{2fG}{3\lambda -1},\;\;\; 
 \tilde{V}(\hat{\chi})\equiv \frac{{V}(\hat{\chi})}{f(\lambda)}.
 \eq
Note that in writing the above equations, we had set the cosmological constant  $\Lambda = 0$. On the other hand,  the Klein-Gordon equation in the flat FRW background reads,
 \bq
 \lb{scalarflat}
\hat{\chi}''+2{\cal{H}}\hat{\chi}'+a^2 \tilde{V}_{,\chi}=0.
\eq
Eqs.(\ref{fried1}), (\ref{fried2}), and (\ref{scalarflat}) are identical to these given in GR \cite{MW09}, if one identifies $\tilde{G}$ and $\tilde{V}$
 to the Newtonian  constant and scalar potential,  respectively. As a result, all the conditions for inflationary models  obtained in GR are equally applicable 
 to the current case, as long as the background is concerned. In particular, the slow-roll conditions,
 \bq
 \lb{slow-roll1}
\tilde \epsilon_{{\scriptscriptstyle V}} ,\; |\tilde\eta_{{\scriptscriptstyle V}}| \ll 1,
 \eq
need to be imposed here, too,  in order to get enough e-fold and solve all the problems encountered in GR,  where
 \bqn
 \lb{slow-roll2}
\tilde \epsilon_{{\scriptscriptstyle V}} &\equiv& \frac{\tilde{M}_{\text{pl}}^{2}}{2}\left(\frac{{\tilde V}'}{\tilde V}\right)^{2} = \frac{3 \lambda -1}{2 f}\epsilon_{{\scriptscriptstyle V}} ,\nb\\
\tilde \eta_{{\scriptscriptstyle V}}  &\equiv& \tilde{M}_{\text{pl}}^{2} \left(\frac{{\tilde V}''}{ \tilde V}\right)= \frac{3 \lambda -1}{2 f}\eta_{{\scriptscriptstyle V}},
 \eqn
with $\tilde{M}_{\text{pl}}^{2} \equiv 1/(8\pi \tilde{G})$, and $\epsilon_{{\scriptscriptstyle V}}$ and $\eta_{{\scriptscriptstyle V}}$ are the ones defined in GR \cite{Inflation}.
However, due to the presence of high-order spatial derivatives, the perturbations will be dramatically different, as to be shown below.

\subsection{Gauge Freedom and Invariants} 

The linear scalar perturbations are given by
\bqn
\lb{pertu}
&& \delta N=a\phi,\;\; \delta N_i=a^2\partial_i B,\;\;
 \delta g_{ij}=-2a^2(\psi\delta_{ij}- \partial_i\partial_jE),\nb\\
&& A=\hat{A}+\delta A,\;\;\varphi=\hat{\varphi}+\delta \varphi,\;\;\chi=\hat{\chi}+\delta \chi,
\eqn
where $\hat{A}$ and $\hat{\varphi}$ denote the background fields of $A$ and $\varphi$, and are functions of $\eta$ only. 
Under the gauge transformations (\ref{1.4}), the scalar perturbations transform as
\bqn
\lb{4.0b}
\tilde{\phi} &=& \phi - {\cal{H}}\xi^{0} - \xi^{0'},\;\;\;
\tilde{\psi} = \psi +  {\cal{H}}\xi^{0},\nb\\
\tilde{B} &=& B +  \xi^{0} - \xi',\;\;\;
\tilde{E} = E -   \xi,\nb\\
\widetilde{\delta\varphi} &=& \delta\varphi - \xi^0 \hat{\varphi}',\;\;\;
\widetilde{\delta{A}} = \delta{A} - \xi^0 \hat{A}' - \xi^{0'} \hat{A}, ~~~
\eqn
where $f = - \xi^0(\eta),\; \zeta^i = - \xi^{,i}(\eta, x)$. Under the $U(1)$ gauge transformations, on the other hand,
we find that
\bqn
\lb{4.0c}
\tilde{\phi} &=& \phi,\;\;\; \tilde{E} = E,\;\;\;
\tilde{\psi} = \psi,\;\;\;
\tilde{B} =  B - \frac{\epsilon}{a}, \nb\\
\widetilde{\delta\varphi} &=& \delta\varphi + \epsilon,\;\;\;
\widetilde{\delta{A}} = \delta{A} - \epsilon',
\eqn
where $\epsilon = - \alpha(\eta, x)$. Then, the  gauge transformations of the whole group $ U(1) \ltimes {\mbox{Diff}}(M, \; {\cal{F}})$
will be the linear combination of the above two. Out of the six unknowns, ($\phi, B, \psi, E, \delta{A}, \delta\varphi$), one can construct four
gauge-invariant quantities,
\bqn
\lb{4.0d}
\gamma&=&\Delta \phi,\;\;\;\;\;\;\nb\\
\Phi &=& \phi - \frac{1}{a - \hat{\varphi}'}\big(a\sigma - \delta\varphi\big)' \nb\\
& & -  \frac{1}{\big(a - \hat{\varphi}'\big)^2}\big(\hat{\varphi}'' - {\cal{H}}\hat{\varphi}'\big)
\big(a\sigma- \delta\varphi\big),\nb\\
\Psi &=& \psi +  \frac{{\cal{H}}}{a - \hat{\varphi}'}\big(a\sigma- \delta\varphi\big),\nb\\
\Gamma &=& \delta{A} +\Bigg[\frac{a\big(\delta\varphi - \hat{\varphi}'\sigma\big) - \hat{A}\big(a\sigma - \delta\varphi\big)}{a - \hat{\varphi}'}\Bigg]',
\eqn
which is different from the projectable case, in which only three of gauge-invariants exist  \cite{HW}, because the projectability condition requires $\phi = \phi(\eta)$, 
which can be eliminated by the gauge transformation $f = - \xi^0(\eta)$. But, now since $\phi = \phi(t, x)$, this becomes impossible. 
Thus, from above discussions and the gauge transformations (\ref{4.0b}) and (\ref{4.0c}), one can choose the following four different 
gauges,
\bqn
\lb{gauges}
&&  (i) \; B=0, \;\delta\varphi=0; \;\;\;\;\;
(ii) \; B=0, \;\delta A=0; \nb\\
&& (iii) \; E=0,\;\delta\varphi=0; \;\;\;
 (iv) \; E=0,\;\delta A=0.
 \eqn

The field equations for the linear scalar perturbations without specifying to any gauge are presented in Appendix A. 
 
\subsection{Equations of Linear Perturbations}

In this subsection,  we consider the cosmological perturbations   with the gauge,
\bq
\lb{gauge}
\hat{\varphi}(\eta)=0,\;\;\;\; E = 0=\delta\varphi.
\eq
Then,  Eqs. (\ref{phichi}) - (\ref{Echi}) can be cast in the following forms,
\bqn
\lb{phi gauge}
&& \frac{3\lambda-1}{2} {\cal{H}}  \Big[3\psi'+3\phi {\cal{H}} +\partial^2 B\Big]-\wp \partial^2 \psi+\frac{1}{2}\eth\partial^2\phi \nb\\
&& ~~~ =-4\pi Ga^2 \Big[\frac{f \hat{\chi}'}{a^{2}} \big(\delta \chi' - \hat{\chi}' \phi\big)  +  \frac{V_4}{a^{4}} \partial^4\delta \chi + {V'} \delta \chi\Big],\nb\\
\\
\lb{A gauge}
&& \partial^2\psi=4\pi Gc_{1} \partial^2 \delta \chi,\;\;\;\;\;\;\;\;\;\;\;\;\;\;\\
\lb{varphi gauge}
&& \partial^2\Big[ (1-\lambda)(3\psi'+\partial^2B)+2 {\cal{H}} \psi+(1-3\lambda) {\cal{H}} \phi \Big]\nb\\
&& ~~~ =8\pi G \Big[\big({c}'_{1} \hat{\chi}'+ {c}_{1}{\cal H} - f \hat{\chi}'\big)\partial^2 \delta \chi + {c}_1 \partial^2 \delta \chi' \Big],\;\;\;\;\;\;\;\;\\
\lb{B gauge}
&& \partial^2\Bigg[(3\lambda-1)(\psi'+\phi {\cal{H}} )+(\lambda-1)\partial^2B\Bigg]\;\;\;\nb\\
&& ~~~~~~~~~~~~~~~~~~~~~~~~~~~~~~  =8\pi G  f \hat{\chi}'\partial^2 \delta \chi, \\
\lb{psi gauge}
&& -\sigma'-2{\cal{H}} \sigma-\psi-\alpha_1\partial^2\psi+\wp\phi\;\;\;\;\;\;\;\nb\\
&& ~~~~~~~~~~~~~~~~~~~~~~~~~~~~~~  +\frac{\hat{A}\psi-\delta A}{a}=0,\\
\lb{E gauge}
&& \psi''+{\cal{H}} (\phi'+2\psi')+({\cal{H}} ^2+2{\cal{H}}')\phi\nb\\
&& +\frac{\lambda-1}{3\lambda-1} \partial^2\Big[\psi+\alpha_1\partial^2\psi-\wp\phi
-\frac{\hat{A}\psi-\delta A}{a}\Big]\nb\\
&& =\frac{8\pi G a^2}{3\lambda-1} \Big[\frac{f \hat{\chi}'}{a^2} (\delta\chi'-\hat{\chi}'\phi)-V'\delta\chi\Big].
\eqn
The Klein-Gordon equation (\ref{1kg}) now reads
\bqn
\lb{1kg2}
&& f \delta\chi'' + 2 {\cal{H}} f \delta\chi' - f \hat{\chi}' (\partial^2B) - 2 {\cal{H}} f \hat{\chi}' \phi \nb\\
&& ~~ - f \hat{\chi}'' \phi - f \hat{\chi}' \phi' + a^2 V'' \delta\chi \nb\\
&& ~~ - 2 \left(\frac{1}{2}+V_1 - \frac{V_2+v_4'}{a^2}\partial^2 - \frac{V_6}{a^4} \partial^4\right) (\partial^2\delta\chi) \nb\\
&&~~  - 3 f \hat{\chi}' \psi' - 2 (c_1'-c_2) \frac{\hat{A}}{a} (\partial^2 \delta\chi) \nb\\
&&~~  - c_1 \partial^2 \left(\frac{\delta A}{a}\right)+ a^2 V' \phi + V_4 (\partial^4\phi)=0.
\eqn

On the other hand, the comoving curvature perturbation, defined by
 \bq
 \lb{6.6}
 {\cal{R}}  = \psi + \frac{\cal{H}}{\bar{\chi}'} \delta\chi,
 \eq
 is  gauge-invariant even in the current setup.  Then, with the help of its definition, and from Eqs.(\ref{phi gauge}), (\ref{A gauge}), (\ref{B gauge}), and (\ref{psi gauge}), 
  we can express $\psi, \; B$ and $ \delta{A}$ in terms of $\delta\chi$, and then submit them into Eq.(\ref{2action}), we obtain the second order action in terms of   $\mathcal{R}$,
\bqn
\lb{actionR}
&& S^{(2)} = \frac{1}{2}\int d\eta d^3 x a^2 h^2\Big[ {\cal{R}}' e_0 {\cal{R}}' + {\cal{R}} e_1 {\cal{R}} \Big],
\eqn
where
\bqn
\lb{6.72b}
h &\equiv& \left(4\pi G c_1 + \frac{H}{\dot{\bar{\chi}}}\right)^{-1} = \frac{\delta \chi}{{\cal{R}}},\nb\\
e_0 &\equiv& f +\frac{4 \pi G c^2_1}{ |c^2_{\psi}|} - \frac{4 \zeta^2d_1^2}{ d_0},\nb\\
e_1 &\equiv& - a^2 V''+ 4 \pi G a^2 V' c_1 \left(\frac{c_1'}{f |c_\psi^2|} - \frac{1}{|c_\psi^2|}+3\right)  \nb\\
&& + \frac{8\pi G f \hat{\chi}'^2}{\lambda-1} (f-c_1') - \frac{4 \pi G c_1 c_1'' \hat{\chi}'^2}{|c_\psi^2|} \nb\\
&&+ \left[1+ 2 V_1 + 2 \frac{\hat{A}}{a} (c_1' - c_2) - 4\pi G c^2_1 (1-\frac{\hat{A}}{a})\right]\partial^2 \nb\\
&& - 2 \left(\frac{V_2 + V_4'}{a^2} + 2\pi G c^2_1 \alpha_1\right)\partial^4 - \frac{2V_6}{a^4}\partial^6 \nb\\
&& - \frac{4 \zeta^2 d_2^2}{ d_0} + \frac{8 \zeta^2 {\cal{H}} d_1 d_2}{ d_0} + 4 \zeta^2 \left(\frac{d_1d_2}{d_0}\right)' \nb\\
&& - \left(\frac{h''}{h}\right) e_0 - \left(\frac{h'}{h}\right) (e_0'+2 {\cal{H}} e_0),
\eqn
with
\bqn
c_\psi^2&\equiv&\frac{\lambda-1}{1-3\lambda},\nb\\
d_0&\equiv& \left(\frac{3\lambda-1}{\lambda-1}\right){\cal{H}}^2 + 4 \pi G f \hat{\chi}'^2 - \frac{1}{2} \eth \partial^2,\nb\\
d_1 &\equiv& 4 \pi G \left[f\hat{\chi}'-\left(\frac{3\lambda-1}{\lambda-1}\right) c_1 {\cal{H}}\right],\nb\\
d_2 &\equiv& 4 \pi G \Bigg[a^2 V' + \left(\frac{3\lambda-1}{\lambda-1}\right) (f \hat{\chi}'-c_1' \hat{\chi}' {\cal{H}})\nb\\
&&\;+ \frac{V_4}{a^2}\partial^4 - c_1 \wp \partial^2\Bigg].
\eqn
From the above expressions, it can be shown that Eq.(\ref{phi gauge}) can be cast in the form,
\bq
\lb{aa}
d_0\phi=d_1\delta\chi'+d_2\delta\chi.
\eq

Introducing the variables, 
\bq
\lb{6.72c}
v \equiv  z {\cal{R}}, \;\;\;
z^2 \equiv a^2 h^2 e_0,
\eq
from Eq.(\ref{actionR}) one obtains the equation of mode function $v_k$ in the momentum space, which reads
\bq
\lb{eom}
v''_k + \beta^2 (\eta,k) v_k =0,
\eq
where
\bqn
\lb{6.72fa}
\beta^2(\eta,k)  &=& - \left(\frac{z''}{z}+\frac{e_1}{e_0}\right)_{\partial_{i} \rightarrow i k_{i}} \nb\\
  &=& \omega^2(\eta,k)+m_{\text{eff}}^2,
\eqn
and the exact expressions of $\beta^2(\eta,k)$, $\omega^2(\eta,k)$, and $m_{\text{eff}}$ are given in Appendix B. It is remarkable to note  that with the U(1) symmetry,
a master equation (\ref{eom}) exists, in contrast to the case without it \cite{BF02}. This is closely related to the fact that the U(1) symmetry eliminates the spin-0 graviton,
and the resulted theory enjoys the same degree of freedom as that of GR. 

\section{Power Spectrum and Index}

To calculate the power spectrum of the linear scalar perturbations with slow-roll approximations, as one usually does   in GR, we
 should first choose a specific vacuum state of the quantum field at initial time $\eta_i$ (the initial conditions) and solve the linearized equation of motion.
 Then, using both the initial conditions   the solution is uniquely determined, from which we  calculate   the power spectrum and index. 
 
 However, due to the complexity of
 the function $\beta^2(\eta,k)$ defined in Eq.(\ref{6.72fa}),   the analysis becomes  very much  involved mathematically. In the following  we  follow \cite{Martin}.  
 First,  we notice that,  at the beginning of the inflationary period, the physical wavelengths 
 of comoving scales which correspond to present large-scale structure of the Universe are usually much smaller than the corresponding Hubble length. 
 This means that at the early stage of inflation the high order spatial derivative terms 
 of the theory  are dominant.  In this case, 
a natural choice of    the initial state is the one  that 
 minimizes the energy   of the field at  the initial moment $\eta_i$ \cite{Martin}, 
\bqn\lb{initial conditions}
v_k(\eta_i) &=& \frac{1}{\sqrt{2\beta(\eta_i,k)}},\nb\\
v_k'(\eta_i) &=& \pm i \sqrt{\frac{\beta(\eta_i,k)}{2}}.
\eqn
Then, with the slow roll approximations,  one has
\bqn
\beta^2(\eta_i,k) \simeq \gamma^2 \eta^4k^6,
\eqn
where
\bqn
\lb{Mstar}
\gamma^2 \equiv - \frac{2 (1-4\epsilon_V)V_6H^4 }{f+4\pi G c_1^2/|c_\psi^2|} \simeq (1-4\epsilon_V)\left(\frac{H}{M_{\star}}\right)^{4}c_s^2. ~~~~~
\eqn
In the above expression, $c_s$ is defined by (\ref{cs}) and $M_{\star} \le M_{\text{pl}}$. For the field to be stable in the UV, the condition $V_6<0$ has to be satisfied, which in the following will be assumed to be always the case.

Secondly, as can be seen from (\ref{beta}), the expression of $\beta^2(\eta,k)$ is too complicated to solve the corresponding equation of motion analytically. 
Following \cite{Martin}, we divide the momentum space into three regions:
 \begin{itemize}
  \item Region I, in which we have $k_{\text{phy}}>M_\star$, and the $k^6$ term of $\beta^2(\eta,k)$ dominates; 
  \item Rgion II, in which we have $M_\star >k_{\text{phy}}>H/c_s$, and the  $k^2$ term in $\beta^2(\eta,k)$ dominates; 
  \item Region III, in which we have $k_{\text{phy}}<H/c_s$, and the  $m_{\text{eff}}$ term   dominates, 
 \end{itemize}
In the following, we will solve the equation of motion (\ref{eom}) in each of these three regions, separately, and then using the matching conditions on the boundaries to
connect the integration constants obtained in each of the three regions. In particular, the coefficients of the two fundamental solutions in Region I are fixed by the initial conditions 
(\ref{initial conditions}). Then, we perform the matching of $v_k$ and $v_k'$ at the transitions between region I and region II, which occurs at time $\eta_1$, and between regions II and III, which occurs at time $\eta_2$, to obtain the coefficients of the two fundamental solutions in region III, from which the power spectrum and index  can be calculated.

\subsection{Region I}

In Region I, the equation of motion for the mode function $v_k$ reduces to
\bqn
v_k''+\gamma^2 \eta^4 k^6 v_k=0,
\eqn
which has the general solution,
\bqn
v_k^I(\eta)=A_1 |\eta|^{1/2} J_\nu(z)+A_2 |\eta|^{1/2} J_{-\nu}(z),
\eqn
where $J_{\nu}(z)$ denotes the usual Bessel function, with
 $\nu=1/6$ and $z=\gamma |\eta|^3k^3$/3. The coefficients $A_1$ and $A_2$ are to be determined by   the initial conditions 
 (\ref{initial conditions})  at $\eta_i$, i.e.,
\bqn
A_1J_\nu(z_i)+A_2J_{-\nu}(z_i)=v_k(\eta_i) |\eta_i|^{-1/2},\nb\\
- A_1 J_{\nu-1}(z_i) + A_2 J_{1-\nu}(z_i)=\frac{v_k'(\eta_i) |\eta_i|^{1/2}}{3 z_i}.
\eqn
Using the Wronskian relation $J_{-\nu}(z)J_{\nu-1}(z)+J_{1-\nu}(z)J_{\nu}(z)=2 \sin[\nu \pi]/(\pi z)$, we find that
\bqn
A_1&=&\pi z_i v_k(\eta_i) |\eta_i|^{-1/2} [J_{1-\nu}(z_i) \mp i J_{-\nu}(z_i)],\nb\\
A_2&=&\pi z_i v_k(\eta_i) |\eta_i|^{-1/2} [J_{\nu-1}(z_i) \pm i J_{\nu}(z_i)].
\eqn
Since $z_i \gg 1$, by introducing two new functions,
\bq
x(\eta) \equiv z(\eta)+\frac{\nu \pi}{2} -\frac{\pi}{4},\;\;y(\eta)\equiv z(\eta)-\frac{\nu \pi}{2} -\frac{\pi}{4},
\eq
and using the asymptotic forms of the Bessel function for large arguments, we find
\bqn\lb{A1}
A_1 \simeq \mp i \sqrt{2 \pi z_i} v_k(\eta_i) |\eta_i|^{-1/2} e^{\pm i x_i},\nb\\
A_2 \simeq \pm i \sqrt{2 \pi z_i} v_k(\eta_i) |\eta_i|^{-1/2} e^{\pm i y_i}.
\eqn

\subsection{Region II}
In Region II, the dispersion relation is approximatively to be linear, thus the equation of motion reduces to
\bqn
v_k''+c_s^2 k^2v_k=0,
\eqn
with
\bqn\lb{cs}
c_s^2 \equiv \frac{1+2V_1+2 \bar{A} (c_1'-c_2)-4\pi G c_1^2 (1-\bar{A}-\frac{2}{\beta_0})}{f+4\pi G c_1^2/|c_\psi^2|} .\nb\\
\eqn
Therefore,  the solution can be expressed as
\bqn
v_k^{II}(\eta) = B_1 e^{i c_s k \eta} +B_2 e^{- i c_s k \eta}.
\eqn
The coefficients $B_1$ and $B_2$ are to be determined by matching of $v_k$ and $v_k'$ at the transitions time $\eta_1$ between regions I and II, 
which yield
\bqn
&& B_1 e^{i c_s k \eta_1} = \frac{A_1|\eta_1|^{1/2} }{2} \left[J_\nu(z_1)-\frac{3 z_1|\eta_1|^{-1}}{i c_s k }  J_{\nu-1 }(z_1)\right]\nb\\
&& ~~~ + \frac{A_2|\eta_1|^{1/2} }{2} \left[J_{-\nu}(z_1)+\frac{3 z_1|\eta_1|^{-1}}{i c_s k }  J_{1-\nu}(z_1)\right],\nb\\
&& B_2 e^{-i c_s k \eta_1} =  \frac{A_1|\eta_1|^{1/2} }{2} \left[J_\nu(z_1)+\frac{3 z_1|\eta_1|^{-1}}{i c_s k }  J_{\nu-1 }(z_1)\right]\nb\\
&& ~~~ + \frac{A_2|\eta_1|^{1/2} }{2} \left[J_{-\nu}(z_1)-\frac{3 z_1|\eta_1|^{-1}}{i c_s k }  J_{1-\nu}(z_1)\right].\nb\\
\eqn
For $z_1\gg 1$, we obtain
\bqn
 B_1 &\simeq &\frac{A_1|\eta_1|^{1/2} e^{-i c_s k \eta_1}}{\sqrt{2\pi z_1}} \left[\cos{y_1}+\frac{3 z_1}{i c_s k |\eta_1|} \sin{y_1}\right]\nb\\
 && + \frac{A_2|\eta_1|^{1/2} e^{-ic_s k \eta_1}}{\sqrt{2\pi z_1}} \left[\cos{x_1}+\frac{3 z_1}{ic_s k |\eta_1|} \sin{x_1}\right],\nb\\
 B_2 &\simeq &\frac{A_1|\eta_1|^{1/2} e^{ic_s k \eta_1}}{\sqrt{2\pi z_1}} \left[\cos{y_1}-\frac{3 z_1}{ic_s k |\eta_1|} \sin{y_1}\right]\nb\\
 && + \frac{A_2|\eta_1|^{1/2} e^{ic_s k \eta_1}}{\sqrt{2\pi z_1}} \left[\cos{x_1}-\frac{3 z_1}{ic_s k |\eta_1|} \sin{x_1}\right],\nb\\
\eqn
where $A_1$ and $A_2$ are given by Eq.(\ref{A1}) and the transitions time $\eta_1$ is given by
\bqn
|\eta_1| = \frac{1+\epsilon_V}{H} \frac{M_\star}{k}.
\eqn

\subsection{Region III and the power spectrum}
In Region III, the equation of motion reads
\bqn
v_k'' + m_{\text{eff}}^2 v_k=0,
\eqn
where $m_{\text{eff}}^2$ is the effective mass term and is  given by Eq.(\ref{mass}). Then,  the solution can be simply expressed as
\bqn
v_k^{III}(\eta) \simeq C |\eta|^{\frac{1-\sqrt{9-4 w}}{2}},
\eqn
where $w=3 \eta_V/f +9 (2 |c_\psi^2|-1) \epsilon_V$, and the coefficient $C$ can be determined by matching $v_k$ and $v_k'$ at  $c_sk=aH$, which gives
\bqn
C=\left(B_1 e^{ic_s k \eta_2}+B_2 e^{-i c_s k \eta_2}\right)|\eta_2|^{\frac{-1+\sqrt{9-4 w}}{2}}.
\eqn
Set Then the power spectrum of comoving curvature perturbation $\mathcal{R}$ can be calculated, and is given by 
\bqn
\frac{k^3}{2\pi^2}P_{\mathcal{R}} &=& \left(\frac{H}{2\pi}\right)^2\left(\frac{1+2\epsilon_V}{c_s^3 h^2e_0}\right)\left|c_sk\eta\right|^{-2w/3}.
\eqn
Setting the slow roll parameters to zero exactly, the power spectrum given above can be put in the simple form,
\bqn
\frac{k^3}{2\pi^2}P_{\mathcal{R}} =\left(\frac{H}{2\pi}\right)^2\left(\frac{1}{c_s^3 h^2e_0}\right)_{c_sk\eta\rightarrow 0},
\eqn
here
\bqn
e_0|_{c_sk\eta\rightarrow 0}&=&f+\frac{4\pi G c_1^2}{|c_\psi^2|}\Bigg\{1 -\frac{1}{1/|c_\psi^2|+\frac{1}{2}\beta_0} \Bigg\}.\nb\\
\eqn
We can see from above that the scale-invariance of the spectrum is maintained in our theory. In the relativistic limit ($c_1=c_2=\beta_0=\bar{A}=V_1=0,\;\;\lambda=1=f$), the power spectrum reduces to the well known result obtained in GR.

The spectrum index can be calculated as
\bqn\lb{index}
n_{\mathcal{R}}-1=2\eta_V-6\epsilon_V+\Delta n_{\mathcal{R}1}+\Delta n_{\mathcal{R}2},
\eqn
where the modification of the index is
\bqn
2\eta_V-6\epsilon_V+\Delta n_{\mathcal{R}1}\equiv- \frac{2}{3}w,
\eqn
which comes from the effective mass term $m_{\text{eft}}$, and
\bqn\lb{index1}
\Delta n_{\mathcal{R}2}\equiv- \frac{d\ln e_0}{d\ln k},
\eqn
which represent the contributions from higher curvature terms in $e_0$. To estimate the effect from higher curvature terms on the index, let us first define
\bqn
\delta \equiv \frac{2|c_\psi^2|}{ \beta_0|c_\psi^2|+2},\;\;
M_{A} \equiv  \frac{M_{pl}}{|\beta_2+\beta_4|^{1/2}},\;\;
M_{B} \equiv  \frac{M_{pl}}{|\beta_8|^{1/4}}.\nb\\
\eqn
Then,  $e_0$ can be expressed as
\bqn
e_0&=& f+\frac{4\pi G c_1^2}{|c_\psi^2|}\nb\\
&& -\frac{4\pi G c_1^2}{1+\frac{1}{2}\beta_0|c_\psi^2|}\frac{1}{1-\delta \epsilon_1|c_sk\eta|^2-\delta\epsilon^2_2|c_sk\eta|^4},\nb\\
\eqn
where
\bqn
\epsilon_1=\frac{H^2}{M_A^2},\;\;\;\epsilon_2=\frac{H^2}{M_B^2}.
\eqn
Usually,  the introduced scales $M_A$ and $M_B$ are much greater than  $H$ when inflation just starts, i.e., $\epsilon_{1,2} \ll 1$. For the sake of simplicity, we assume that $c_1 \simeq c_2 \simeq M_\star$
\cite{HWWb}.
Then, we  have
\bqn\lb{index2}
\frac{d\ln e_0}{d\ln k} &\simeq&- \frac{1}{e_0(1+\frac{1}{2}\beta_0|c_\psi^2|)}\left(\frac{M_\star}{M_{\text{pl}}}\right)^2\nb\\
&& \times (\delta\epsilon_1+2\delta\epsilon_1^2+2 \delta \epsilon_2^2).
\eqn

Two comments are in order. First, it appears that the nonlinear dispersion relation $\beta^2(\eta,k)$ has insignificant contributions to the standard spectrum $k^3P_{\mathcal{R}}^{\text{GR}}$. This is consistent with the results obtained in Refs.\cite{Martin2,NP},  when two conditions are satisfied. The first one is the adiabaticity that is satisfied for the mode propagation before horizon crossing, i.e., $\mathcal{C}\equiv |\beta'(\eta,k)/\beta^2{\eta,k}| \ll 1$. As shown in Refs.\cite{Martin2,NP}, only when this condition is violated, can the nonlinear dispersion relation have significant contributions to the spectrum. In this paper, we assumed that the adiabatic condition $\mathcal{C}\ll 1$ is always satisfied before horizon crossing. The second condition is that $\beta^2(\eta,k)$ is approximately linear in region II, thus when we evaluate  the power spectrum at horizon crossing, all the high order derivative terms are neglected. More accurately, one can expand $k^3/\beta^3(\eta,k)$ in order of $\epsilon_{\text{HL}}$ in the form, 
\bqn
\left.\frac{k^3}{\beta^3(\eta,k)}\right|_{\beta(\eta_3,k)=aH} \simeq \frac{1}{c^3_s} (1+a_1 \epsilon_{\text{HL}}+a_2 \epsilon^2_{\text{HL}}+\dots),\nb\\
\eqn
where $\epsilon_{\text{HL}} \simeq H^2/M_{\star}$, and $M_{A} \simeq M_{B} \simeq M_{\star}$. Thus,  the power spectrum which contains contributions from $\beta^2(\eta,k)$ can be expressed as
\bqn
\frac{k^3}{2\pi^2}P_{\mathcal{R}} =\left(\frac{H}{2\pi}\right)^2\left(\frac{1}{c_s^3 h^2e_0}\right)(1+a_1 \epsilon_{\text{HL}}+a_2 \epsilon^2_{\text{HL}}+\dots).\nb\\
\eqn
 
Second,  for the spectrum index acquires corrections from both the effective mass term $m_{\text{eff}}$ and $1/e_0$. As shown in (\ref{index2}),  corrections from $1/e_0$ are of order $\epsilon_{\text{HL}}$. From the first term $\mathcal{R}'e_0(\eta,\partial^{2n})\mathcal{R}'$ in the second order action (\ref{actionR}), one can see that these corrections come from the mixture of spatial derivative and time derivative of the comoving curvature perturbations. And more interesting, such mixture is a consequence of the fact that the lapse function $\phi$ is a function of both $\eta$ and $x$, i.e., without the projectability condition. Thus, such corrections to the spectrum index presents a distinguishable signature of the model from the version with the projecttability condition.

\section{Conclusions and Discussions}

In this paper, we have studied inflation driven by a single scalar field in the general covariant Ho\v{r}ava-Lifshitz gravity without projectability condition, formulated in \cite{ZWWS,ZSWW}. Because of the enlarged symmetry (\ref{symmetry}),  
the gauge invariants for cosmological perturbations are different from those given in GR. They are also different from 
  those given in general covariant Ho\v{r}ava-Lifshitz theory with projectability conditions. By using the gauge transformations for perturbations, we have constructed all the four gauge invariants, given by
  Eq.(\ref{4.0d}).

The most general cosmological linear perturbation equations have been derived without specifying any gauge, and presented in Appendix A. By using these equations, we have shown that a master equation for the mode function  of  the scalar perturbations exists, in contrast to the case without the $U(1)$ symmetry\cite{BF02}. Following the method developed in Ref.\cite{Martin}, we have first divided the momentum space into three different regions, and then solved the equation of motion in each of the three regions, separately. With the initial conditions that the vacuum state minimizes the energy of the field, we have determined the coefficients of the solutions by matching the solutions and its time derivative at the initial time and transition times on the boundaries of the three regions. 
 
With these solutions, we have also calculated the power spectra and spectrum index of the scalar perturbations in the slow-roll approximations.  Comparing with the standard results obtained in GR, the power spectrum and index acquire tiny corrections from the theory, as the adiabatic condition holds in the case considered here. Remarkably,  partly of corrections are of the consequence of  the non-projectability condition, i.e., $N=N(t,x)$. In the relativistic limit, the spectrum and index reduce to the standard results given  in GR. It is also true that the scale invariance of the comoving curvature perturbation remains almost the same as that given in GR, although the initial conditions chosen here are very different from those in GR. Thus, the general covariant Ho\v{r}ava-Lifshitz gravity without the projectability condition is consistent with all current observations.

~\\{\bf Acknowledgements:}
This work was supported in part by DOE  Grant, DE-FG02-10ER41692 (AW);  NSFC No. 11173021 (AW);  NSFC No. 11075141 (AW); NSFC No. 11105120  (TZ); and NSFC No. 11047008 (TZ).

\section*{Appendix A:  Cosmological Perturbations}
\renewcommand{\theequation}{A.\arabic{equation}} \setcounter{equation}{0}

In this section, we shall consider the linear perturbations given by Eq.(\ref{pertu})
in a flat FRW universe, $k = 0$.  Then, it can be shone that the action to second-order of ($\phi, B, \psi, E, \delta{A}, \delta\varphi$)
takes the form, 
\bqn
\lb{2action}
S^{(2)}&=&\zeta^2 \int d\eta d^3x \Big[\delta_2 (\sqrt{g} N {\cal{L}}_K)-\delta_2(\sqrt{g}N{\cal{L}}_V)\nb\\
&+&\delta_2(\sqrt{g}N{\cal{L}_A})+\delta_2(\sqrt{g}N{\cal{L}}_\varphi)+\frac{\delta_2(\sqrt{g}N{\cal{L}}_\chi)}{\zeta^2}\Big],\nb\\
\eqn
where
\bqn
\lb{2LK}
\delta_2 (\sqrt{g} N {\cal{L}}_K)&=& (1-3\lambda) a^2 \Big[ 6 {\cal{H}} (\phi+\psi) \psi'+3 \psi'^2  \nb\\
&& +  3 {\cal{H}}^2  \phi^2 + \frac{9}{2} {\cal{H}}^2 \psi^2 + 9 {\cal{H}}^2 \phi\psi \nb\\
&& - 2 {\cal{H}} (\phi+\psi) (\partial^2\sigma) - 2 \psi' (\partial^2\sigma) \nb\\
&& - 2 {\cal{H}} \psi' (\partial^2E) - 2 {\cal{H}} (\partial^2E) (\partial^2\sigma) \nb\\
&& - 2 {\cal{H}} \psi (\partial^2B) - 2 {\cal{H}} (\partial^2E)(\partial^2B) \nb\\
&& - 3 {\cal{H}}^2 (\phi+\psi) (\partial^2E) - \frac{3}{2} {\cal{H}}^2 (\partial^2E)^2\Big]\nb\\
&& + (1-\lambda) a^2 (\partial^2\sigma)^2,\\
\lb{2LV}
\delta_2 (\sqrt{g} N {\cal{L}}_V)&=& a^4 \Lambda \Big[ - 6 \phi \psi + 2 (\phi-\psi) (\partial^2E) \nb\\
&&\;\;\;\;\;\;\;\;\;+ 3 \psi^2 - (\partial^2E)^2 \Big] \nb\\
&& + a^2 \Big[  2 \psi (\partial^2\psi)  + 2 \alpha_1 \psi (\partial^4 \psi) \nb\\
&&\;\;\;+ \phi (\eth \partial^2 \phi) - 4 \phi (\wp \partial^2\psi) \Big],\\
\lb{2LA}
\delta_2 (\sqrt{g} N {\cal{L}}_A)&=& a^4 \Lambda_g \Big[ \frac{2 \delta A}{a}\left(- 3 \psi+\partial^2 E\right)\nb\\
&&+\frac{\hat{A}}{a}\left(3\psi^2-2\psi\partial^2E-(\partial^2E)^2\right)\Big]\nb\\
&& + 2 a^2 \Big[\frac{\hat{A}}{a} \psi \partial^2 \psi-\frac{2 \delta A}{a}\partial^2\psi\Big],\\
\lb{Lvarphi}
\delta_2 (\sqrt{g} N {\cal{L}}_\varphi)&=& 2 a^3 \Lambda_g \Bigg[ \hat{\varphi} \Big[ \psi (\partial^2 B) +(\partial^2E)(\partial^2B) \nb\\
&& +\psi(\partial^2\sigma)-3 \psi \psi' + \psi' (\partial^2 E) \nb\\
&& + (\partial^2E)(\partial^2\sigma)-\frac{9}{2} {\cal{H}} \psi^2 \nb\\
&& + 3 {\cal{H}} \psi (\partial^2 E) + \frac{3}{2} {\cal{H}} (\partial^2 E)^2\Big]\nb\\
&& + \delta \varphi \Big[3 \psi' - \partial^2 \sigma + 3 {\cal{H}} (3 \psi-\partial^2E)\Big]\Bigg]\nb\\
&& + a^2 \Lambda_g \delta \varphi (\partial^2 \delta \varphi)\nb\\
&& + a \Bigg[ - 2 \hat{\varphi} [{\cal{H}} \psi (\partial^2 \psi) + 2 \psi' (\partial^2 \psi)]\nb\\
&&+4 {\cal{H}} (\psi-\phi) (\partial^2 \delta \varphi) \Bigg]\nb\\
&&+2a (1-\lambda)\Bigg [3 {\cal{H}} \phi (\partial^2\delta \varphi)+3\psi' (\partial^2 \delta \varphi)\nb\\
&& - (\partial^2\delta \varphi) (\partial^2\sigma)\Bigg]\nb\\
&&+ (1-\lambda) (\partial^2 \delta \varphi)^2,\nb\\
\lb{Lchi}
\delta_2(\sqrt{g}N{\cal{L}}_\chi)&=& \frac{a^2 f}{2}\big[\delta\chi'^2+2\hat{\chi}' \delta\chi  (\partial^2B)\nb\\
&&\;\;\;\;\;\;\;\;-4 \hat{\chi}' \phi \delta\chi'+3 \hat{\chi}'^2\phi^2\big]\nb\\
&& - \frac{1}{2} a^4 V'' \delta\chi^2 + \left(\frac{1}{2}+V_1\right)a^2 \delta\chi (\partial^2\chi) \nb\\
&& - (V_2+V_4')\delta\chi (\partial^4 \delta\chi) - \frac{V_6}{a^2} \delta\chi (\partial^6 \delta\chi) \nb\\
&& + a (\hat{A}+\hat{\varphi}') (c_1'-c_2) \delta\chi (\partial^2\delta\chi)\nb\\
&& + a c_1 (\delta A + \delta\varphi') (\partial^2\delta\chi) + a f \hat{\chi}' \delta\varphi (\partial^2\chi) \nb\\
&& + (\phi-3\psi+\partial^2E) \Big[a^2 f (\hat{\chi}' \delta\chi' - \phi \hat{\chi}'^2) \nb\\
&&-a^4 V' \delta\chi \Big] - V_4 \phi (\partial^4\delta\chi) \nb\\
&& + \left(\frac{a^2 f}{2} \hat{\chi}'^2 - a^4 V\right) \Big[\frac{3}{2} \psi^2 - \psi (\partial^2E) \nb\\
&& - \frac{1}{2} (\partial^2E)^2 + \phi (- 3 \psi+\partial^2E)\Big].
 \eqn
In writing the above expressions, we have used the following definitions
\bqn
\sigma &\equiv& E'-B,\nb\\
\alpha_1 &\equiv& \frac{8\gamma_2+3\gamma_3}{a^2\zeta^2},\nb\\
\eth &\equiv& \beta_0+\frac{\beta_2+\beta_4}{a^2\zeta^2}\partial^2-\frac{\beta_8}{a^4\zeta^4}\partial^4,\nb\\
\wp &\equiv& 1-\frac{\beta_7}{a^2 \zeta^2}\partial^2.
\eqn

Then Variation of the action (\ref{2action}) with respect to $\phi$ , $\psi$, $B$, $E$, $\delta A$, and $\delta \varphi$ yield, respectively
\bqn
\lb{phichi}
&& \frac{3\lambda-1}{2} {\cal{H}}  \Big[3\psi'+3\phi {\cal{H}} -\partial^2 \sigma+\frac{1}{a}\partial^2\delta\varphi\Big]-\wp \partial^2 \psi+\frac{1}{2}\eth\partial^2\phi \nb\\
&& ~~~ =-4\pi Ga^2 \Big[\frac{f \hat{\chi}'}{a^{2}} \big(\delta \chi' - \hat{\chi}' \phi\big)  +  \frac{V_4}{a^{4}} \partial^4\delta \chi + {V'} \delta \chi\Big],~~~ ~\\
\lb{Achi}
&& \partial^2\psi=4\pi Gc_{1} \partial^2 \delta \chi,\;\;\;\;\;\;\;\;\;\;\;\;\;\;\\
\lb{varphichi}
&& \partial^2\Big[ (1-\lambda)(3\psi'-\partial^2\sigma+\frac{\partial^2\delta\varphi}{a})+2 {\cal{H}} \psi+(1-3\lambda) {\cal{H}} \phi \Big]\nb\\
&& ~~~ =8\pi G \Big[\big({c}'_{1} \hat{\chi}'+ {c}_{1}{\cal H} - f \hat{\chi}'\big)\partial^2 \delta \chi + {c}_1 \partial^2 \delta \chi' \Big],\;\;\;\;\;\;\;\;\\
\lb{Bchi}
&& \partial^2\Bigg[(3\lambda-1)(\psi'+\phi {\cal{H}} )-(\lambda-1)\left(\partial^2\sigma-\frac{\partial^2\delta\varphi}{a}\right)\Bigg]\;\;\;\nb\\
&& ~~~~~~~~~~~~~~ =8\pi G  f \hat{\chi}'\partial^2 \delta \chi,\;\;\;\;\;\;\;\;\;\;\;\;\;\;\;\\
\lb{psichi}
&& -\sigma'-2{\cal{H}} \sigma-\psi-\alpha_1\partial^2\psi+\wp\phi\;\;\;\;\;\;\;\nb\\
&&  ~~~~~~~~~~~~~~ +\frac{\hat{A}\psi-\delta A+\hat{\varphi'}\psi+{\cal{H}} \delta\varphi}{a}=0,\;\;\;\;\\
\lb{Echi}
&& \psi''+{\cal{H}} (\phi'+2\psi')+({\cal{H}} ^2+2{\cal{H}}')\phi\nb\\
&& ~~~~~~~~~~~~~~  +\frac{\lambda-1}{3\lambda-1} \partial^2\Big[\psi+\alpha_1\partial^2\psi-\wp\phi\nb\\
&& -\frac{\hat{A}\psi-\delta A+\hat{\varphi'}\psi+2{\cal{H}} \delta\varphi+\delta\varphi'}{a}\Big]\nb\\
&&  ~~~~~~~~~~~~~~ =\frac{8\pi G a^2}{3\lambda-1} \Big[\frac{f \hat{\chi}'}{a^2} (\delta\chi'-\hat{\chi}'\phi)-V'\delta\chi\Big],
\eqn
and variation with respect to $\delta\chi$, one can get the linear Klein-Gordon equation
\bqn
\lb{1kg}
&& f \delta\chi'' + 2 {\cal{H}} f \delta\chi' + f \hat{\chi}' (\partial^2\sigma) - 2 {\cal{H}} f \hat{\chi}' \phi \nb\\
&& ~~~~~~~  - f \hat{\chi}'' \phi - f \hat{\chi}' \phi' + a^2 V'' \delta\chi \nb\\
&&   ~~~~~~~ - 2 \left(\frac{1}{2}+V_1 - \frac{V_2+v_4'}{a^2}\partial^2 - \frac{V_6}{a^4} \partial^4\right) (\partial^2\delta\chi) \nb\\
&&   ~~~~~~~ - 3 f \hat{\chi}' \psi' - 2 (c_1'-c_2) \frac{\hat{A}+\hat{\varphi}'}{a} (\partial^2 \delta\chi) \nb\\
&&  ~~~~~~~ - c_1 \partial^2 \left(\frac{\delta A+\delta\varphi'}{a}\right) - f \hat{\chi}' \partial^2 \left(\frac{\delta\varphi}{a}\right) \nb\\
&&   ~~~~~~~ + a^2 V' \phi + V_4 (\partial^4\phi)=0.
\eqn

\section*{Appendix B:  $\beta^2(\eta,k)$, $\omega^2(\eta,k)$, and $m_{\text{eff}}(\eta) $}
\renewcommand{\theequation}{B.\arabic{equation}} \setcounter{equation}{0}

The functions $\beta^2(\eta,k)$, $\omega^2(\eta,k)$, and $m_{\text{eff}}$, defined in Eq.(\ref{6.72fa}) are given by,
\bqn\lb{beta}
\beta^2 &=& - {\cal{H}}^2 - {\cal{H}}' - {\cal{H}} \frac{e_0'(k)}{e_0(k)}   -  \frac{1}{2} \frac{e_0''(k)}{e_0(k)} + \frac{1}{4}\left( \frac{e_0'(k)}{e_0(k)} \right)^2 \nb\\
&& - \frac{1}{e_0(k)}  \Bigg\{4 \pi G a^2 V' c_1 \left(\frac{c_1'}{f |c_\psi^2|} - \frac{1}{|c_\psi^2|}+3\right)\nb\\
&& - a^2 V''  + \frac{8 \pi G f \hat{\chi}'^2}{\lambda-1} (f-c_1') - \frac{4 \pi G c_1 c_1'' \hat{\chi}'^2}{|c_\psi^2|} \nb\\
&& - \Bigg[1+2V_1+2 \left(\frac{\hat{A}}{a}\right)(c_1'-c_2)  \nb\\
&& ~~~~ - 4 \pi G c_1^2 (1-\frac{\hat{A}}{a})\Bigg]k^2\nb\\
&& - 2 \left(\frac{V_2+V_4'}{a^2}+2 \pi G c_1^2 \alpha_1 \right)k^4 + \frac{2 V_6}{a^4} k^6 \nb\\
&& - 4 \zeta^2\frac{d_2^2(k)}{d_0(k)} + 8\zeta^2 {\cal{H}}\frac{d_1d_2(k)}{d_0(k)} \nb\\
&&  + 4 \zeta^2 \left(\frac{d_1d_2(k)}{d_0(k)}\right)'\Bigg\},\\
m^2_{\text{eff}} &=& \beta^2 (\eta,0),
\eqn
where
\bqn
 \omega^2 (\eta,k) &\equiv& \beta^2(\eta,k)- m^2_{\text{eff}}(\eta),\nb\\
e_0(k) &=& f + \frac{4 \pi G c_1^2}{|c_\psi^2|} - 4 \zeta^2 \frac{d_1^2}{d_0(k)},\nb\\
d_0(k) &=& \frac{{\cal{H}}^2}{|c_\psi^2|} + 4 \pi G f \hat{\chi}'^2 + \frac{1}{2}\beta_0 k^2 \nb\\
            && - \frac{\beta_2+\beta_4}{2a^2 \zeta^2} k^4 - \frac{\beta_8}{2 a^4 \zeta^4}k^6 \nb\\
d_2(k) &=& 4 \pi G \Bigg[a^2 V' + \frac{(f\hat{\chi'}-c_1' \hat{\chi}'){\cal{H}}}{|c_\psi^2|} + \frac{V_4}{a^2} k^4\nb\\
          && + c_1 k^2 + \frac{c_1 \beta_7}{a^2 \zeta^2} k^4\Bigg].
\eqn
Upto first-order of the slow-roll parameters, the mass $m_{\text{eff}}$ can be simplified to
\bqn\lb{mass}
m^2_{\text{eff}} \simeq - \frac{2-3 \eta_V + 3 \epsilon_V}{\eta^2} - \frac{1}{\eta^2} \Big[3 (1- \frac{1}{f})\eta_V + \frac{12 }{3\lambda-1} \epsilon_V\Big].\nb\\
\eqn

\end{document}